\documentclass[aps,twocolumn,pra,superscriptaddress,amsmath,amssymb]{revtex4-1}

\newcommand{\bras}[1]{\langle#1\rvert}
\newcommand{\kets}[1]{\lvert#1\rangle}

\newcommand{\means}[1]{\langle#1\rangle}

\ifx\pdfoutput\undefined
\usepackage[dvipdfmx]{graphicx}
\usepackage[dvipdfmx]{hyperref}
\else
\usepackage{graphicx}
\usepackage{hyperref}
\fi

\usepackage{txfonts}
\usepackage[T1]{fontenc}
\usepackage{xspace}

\usepackage{dcolumn}
\usepackage{bm}
\usepackage{amsmath}
\usepackage{booktabs}
\usepackage[usenames]{color}
\usepackage{xcolor}

\begin{document}
\let\emph\textit

\title{
Strong enhancement of magnetic susceptibility induced by spin-nematic fluctuations in an excitonic insulating system with spin-orbit coupling
}

\author{Joji Nasu}
\affiliation{
  Department of Physics, Yokohama National University,
  Hodogaya, Yokohama 240-8501, Japan
}
\author{Makoto Naka}
\affiliation{
  Waseda Institute for Advanced Study, Waseda University, Tokyo 169-8050, Japan
}
\author{Sumio Ishihara}
\affiliation{
  Department of Physics, Tohoku University, Sendai 980-8578, Japan
}

\date{\today}
\begin{abstract}
Effects of the spin-orbit coupling (SOC) and magnetic field on excitonic insulating (EI) states are investigated.
We introduce the two-orbital Hubbard model with the crystalline field splitting, which is a minimal model for discussing the exciton condensation in strongly correlated electron systems, and analyze its effective Hamiltonian in the strong correlation limit by using the mean-field theory.
In the absence of the SOC and magnetic field, the ground state changes from the nonmagnetic band-insulating state to the EI state by increasing the Hund coupling.
In an applied magnetic field, the magnetic moment appears in the EI state, which is continuously connected to the forced ferromagnetic state.
On the other hand, in the presence of the SOC, they are separated by a phase boundary.
We find that the magnetic susceptibility is strongly enhanced in the EI phase near the boundary with a small SOC.
This peculiar behavior is attributed to the low-energy fluctuation of the spin nematicity inherent in the high-spin local state stabilized by the Hund coupling.
The present study not only reveals the impact of the SOC for the EI state but also sheds light on the role of quantum fluctuations of the spin nematicity for the EI state.

\end{abstract}

\maketitle

\section{Introduction}

Spontaneous hybridization between valance and conduction bands associated with a symmetry breaking, which is known as the exciton condensation, has been a long-standing subject of condensed matter physics for close to half a century~\cite{mott1961transition,PhysRev.158.462,RevModPhys.40.755,PhysRevB.62.2346,PhysRevLett.89.166403,0953-8984-27-33-333201,doi:10.1143/JPSJ.31.730,doi:10.1143/JPSJ.31.812,doi:10.1143/JPSJ.44.1759}.
This is understood by a pair condensation of a particle and a hole, which was proposed as an analogy of superconductivity.
One of the characteristics of the excitonic states is a deformation of the electronic energy bands around the Fermi level originating from the spontaneous hybridization.
This was recently observed in the layered chalcogenide Ta$_2$NiSe$_5$ using the angle-resolved photoemission spectroscopy, and therefore, it has been intensively studied as a candidate material of the excitonic insulator~\cite{PhysRevLett.103.026402,Wakisaka2012,PhysRevB.87.035121,PhysRevB.90.245144}.
However, unlike the superconducting states exhibiting the Meissner effect, the excitonic insulating (EI) state remains elusive as clear experimental signatures offering conclusive evidence have not been identified yet.

On the other hand, another playground of the EI state was proposed in cobaltites typified by LaCoO$_3$.
These have been studied for decades to clarify the physics of the spin-state transition in strongly correlated electron systems~\cite{PhysRevB.58.R1699,Imada1998,Ravindran1999,Ishikawa2004,Baier2005,Rondinelli2009} while the EI state was originally introduced in the weak coupling regime.
In the cobaltites, due to the competition between the crystalline electric field and Hund coupling, a magnitude of the local spin is changed, e.g., by varying temperature and pressure, between the low-spin (LS) with the total spin $S=0$ for the $t_{2g}^6$ configuration, intermediate-spin (IS) with $S=1$ for $t_{2g}^5 e_g^1$, and high-spin (HS) states with $S=2$ for $t_{2g}^4 e_g^2$ in the Co$^{3+}$ ion.
Thus far, the spin-state transition or crossover has been discussed on the basis of the thermally-mixed spin states.

Meanwhile, in the vicinity of the spin-state transition, distinct local spin states are energetically close to each other, and therefore, the quantum hybridization between them is expected to occur spontaneously.
When this hybridized state coherently appears over an entire crystal, it is regarded as an emergence of the exciton condensation.
Indeed, in the two-orbital Hubbard model, which is the minimal model to describe the essence of the spin-state transition~\cite{PhysRevLett.99.126405}, the emergence of the exciton condensation was suggested between the LS band insulator and HS Mott insulator with the antiferromagnetic (AFM) order due to the interorbital Coulomb interaction~\cite{PhysRevB.85.165135}.
Beyond the model calculations, the first-principles calculation study for Pr$_{0.5}$Ca$_{0.5}$CoO$_3$, which exhibits a characteristic phase transition at $T_s\sim$90K without the change of the space group and the appearance of a magnetic order~\cite{PhysRevB.66.052418,doi:10.1143/JPSJ.73.1987,hejtmanek2013phase}, pointed out a possibility of the EI state~\cite{PhysRevB.90.235112,PhysRevB.89.115134,Sotnikov2017}.
This work and following studies also suggested that the exciton condensation occurs as a kind of multipole orders in the Co ion~\cite{Nasu2016EI,Kaneko2016multipole}.
These stimulate further investigations of the EI states in strongly correlated electron systems~\cite{Sugimoto2018,Li2019excitonic}.

The spin-state transition should occur by applying a magnetic field, and hence, the field-induced exciton condensation is expected in the cobaltites.
Utilizing the modern high magnetic-field measurement, the magnetic properties in LaCoO$_3$ were investigated and several new phases were found under the high magnetic field $\sim 60~{\rm T}$~\cite{Ikeda2016}.
As the candidates of the phases, spin-state orders and EI states are proposed theoretically.~\cite{Altarawneh2012,Tatsuno2016,sotnikov2016field}.

The magnetic-field effect causes the large change of the electronic state due to the strong competition between the energies of distinct spin states.
This is also expected to be brought about by the spin-orbit coupling (SOC), which inevitably exists in transition metal ions with the orbital degeneracy.
The Co$^{3+}$ ion possesses the $t_{2g}$ orbital degree of freedom with an effective angular momentum in the cases of the HS and IS states~\cite{kanamori1957theory1,kanamori1957theory2,Tomiyasu2011}.
Indeed, it was pointed out that the SOC plays a crucial role for a large orbital moment in LaCoO$_3$ by the soft X-ray absorption spectroscopy and the magnetic circular dichroism~\cite{Haverkort2006,Tomiyasu2017Co}.
The SOC yields the mixing between the real orbitals split by the crystalline electric field and will compete or cooperate with the exciton condensation in the vicinity of the spin-state transition/crossover.
However, it remains unclear  how the SOC affects the magnetic properties of the EI state.

In this paper, we study the effects of the SOC and magnetic field on the EI state with strong electron correlations.
From the two-orbital Hubbard model with the crystalline field, we derive an effective Hamiltonian in the strong coupling limit by the perturbation expansion and the effective SOC in the low-energy subspace to reproduce the SOC in the cobaltites.
By applying the mean-field (MF) approximation, we examine the effective model in the vicinity of the phase boundary of the LS and EI phases.
In the absence of the SOC, a phase transition occurs from the LS state to the EI state by applying the magnetic field.
Further increase of the magnetic field causes the continuous change from the EI state to the forced ferromagnetic (FM) one.
When the SOC is introduced, these two states are separated by a phase transition.
We find that the SOC enhances the magnetization induced by an applied magnetic field in the EI state.
This leads to the enhancement of the magnetic susceptibility near the phase boundary between the LS and EI phases.
By analyzing the wavefunction in detail, we reveal that the spin nematicity inherent in the HS $S=1$ states plays a crucial role on the enhancement of the magnetization.
The present results provide a possible route to identify the EI state in experiments under the high magnetic field.

This paper is organized as follows.
In Sec.~\ref{sec:model-method}, we introduce the low-energy effective model to address the exciton condensation and SOC on an equal footing.
The MF theory applied to this model is also presented.
The numerical calculation results are shown in \ref{sec:result}.
First, the results without the SOC is presented in Sec.~\ref{sec:wo-so}, and then, results in the presence of the SOC are shown in Sec.~\ref{sec:with-so}.
The magnetic susceptibility as a function of the Hund coupling as well as the temperature are shown in Sec.~\ref{sec:suscep}.
The origin of the characteristic behavior of the susceptibility shown in the above sections is discussed in Sec.~\ref{sec:suscep}.
Section~\ref{sec:discussion} is devoted to the discussion and summary.

\section{Model and method}\label{sec:model-method}

In the present study, we consider the two orbital Hubbard model with the energy splitting between the orbitals, where the average of the electron number per site is fixed to 2, which corresponds to the half filling.
This is the minimal model to describe the LS and HS states and the transition between them.
In the following, we introduce the Hamiltonian and the effective model in the strong coupling limit on the basis of our previous study~\cite{Nasu2016EI}.
We start from the two orbital Hubbard model ${\cal H}_{\rm Hubbard}={\cal H}_t+{\cal H}_U$, which is given by the local interaction term
\begin{align}
 {\cal H}_U=\Delta\sum_{i\sigma} n_{i a \sigma} +U\sum_{i \eta}n_{i\eta\uparrow}n_{i\eta\downarrow}+U'\sum_{i} n_{ia}n_{ib}\notag\\
 +J\sum_{i\sigma\sigma'}c_{ia\sigma}^\dagger c_{ib\sigma'}^\dagger c_{ia\sigma'}c_{ib\sigma}
 +I\sum_{i\eta\neq \eta'}c_{i\eta\uparrow}^\dagger c_{i\eta \downarrow}^\dagger c_{i\eta'\downarrow}c_{i\eta'\uparrow},\label{eq:9}
\end{align}
and inter-site electron transfer term
\begin{align}
 {\cal H}_t=-\sum_{\means{ij}\eta\sigma}t_\eta (c_{i\eta \sigma}^\dagger c_{j\eta \sigma}+{\rm H.c.}),\label{eq:Ht}
\end{align}
where $c_{i\eta\sigma}$ is the annihilation operator of the electron for orbital $\eta(=a,b)$ with spin $\sigma(=\uparrow,\downarrow)$ at site $i$ and $n_{i\eta\sigma}=c_{i\eta\sigma}^\dagger c_{i\eta\sigma}$ is the number operator.
The positive parameters $\Delta$, $U$, $U'$, $J$, and $I$ represent the crystalline field splitting, the intra- and inter-orbital Coulomb interactions, the Hund coupling, and the pair-hopping interaction, respectively.
In ${\cal H}_t$, we consider the transfer integral $t_\eta$ for the orbital $\eta$ between the nearest neighbor (NN) sites $\means{ij}$.

From the two-orbital Hubbard model in Eq.~(\ref{eq:9}), we derive the effective Hamiltonian in the strong coupling limit.
The low-energy subspace is composed of the direct products of the local electronic states occupied by two electrons, which are the eigenstates of ${\cal H}_U$.
At each site, we consider the following four states: The spin-singlet state $\kets{L}$ for the LS one with $S=0$, and three spin-triplet states $\kets{\Gamma}$ ($\Gamma=X,Y,Z$) for HS ones with $S=1$~\cite{Nasu2016EI}.
The local wavefunctions are explicitly given by
\begin{align}
 \kets{L}&=\left(f c_{b\uparrow}^\dagger c_{b\downarrow}^\dagger - g c_{a\uparrow}^\dagger c_{a\downarrow}^\dagger\right)\kets{0},\label{eq:5}\\
 \kets{X}&=\frac{1}{\sqrt{2}}\left(-c_{a\uparrow}^\dagger c_{b\uparrow}^\dagger+c_{a\downarrow}^\dagger c_{b\downarrow}^\dagger\right)\kets{0},\\
 \kets{Y}&=\frac{i}{\sqrt{2}}\left(c_{a\uparrow}^\dagger c_{b\uparrow}^\dagger+c_{a\downarrow}^\dagger c_{b\downarrow}^\dagger\right)\kets{0},\\
 \kets{Z}&=\frac{1}{\sqrt{2}}\left(c_{a\uparrow}^\dagger c_{b\downarrow}^\dagger + c_{a\downarrow}^\dagger c_{b\uparrow}^\dagger\right)\kets{0},
\end{align}
where $f=\left[1+\left(\Delta-\Delta'\right)^2/I^2\right]^{-1/2}$ and $g=\sqrt{1-f^2}$ with $\Delta'=\sqrt{\Delta^2+I^2}$.
In the HS states, each orbital is occupied by one electron.
On the other hand, the weight of the two-electron occupied state in $\kets{L}$ exists mostly in the $b$ orbital in the case of $g\ll 1$ with the small pair-hopping interaction. 
Note that the three HS states $\kets{X}$, $\kets{Y}$, and $\kets{Z}$ are known as the bases of the spin-nematic states for spin $S=1$~\cite{Chen1971Quadrupole,Chandra1991nematics,Lauchli2006Quadrupolar,Shannon2006Nematic,Tsunetsugu2006Nematic}.
The nematic state $\kets{\Gamma}$ is characterized by a rod-like director along the $\Gamma$ axis in the spin space (see Fig.~\ref{fig_nematic}).

By applying the second-order perturbation expansion with respect to ${\cal H}_t$, the effective Hamiltonian is obtained as~\cite{PhysRevB.89.115134,PhysRevLett.107.167403,PhysRevB.86.045137,Nasu2016EI}
\begin{align}
  {\cal H}_{\rm el}^{\rm eff}=-\tilde{\Delta}\sum_i\tau_i^z+J_{z}\sum_{\means{ij}}\tau_i^z\tau_j^z
+J_s\sum_{\means{ij}}\bm{S}_i\cdot\bm{S}_j\notag\\
-J_x\sum_{\means{ij}\Gamma}\tau_{i\Gamma}^x\tau_{i\Gamma}^x
-J_y\sum_{\means{ij}\Gamma}\tau_{i\Gamma}^y\tau_{j\Gamma}^y,\label{eq:1}
\end{align}
where the constant terms are omitted.
In addition to the spin operators $\{S_i^X,S_i^Y,S_i^Z\}$ for the $S=1$ triplet states, we introduce the pseudospin (PS) operators, $\tau_\Gamma^x$, $\tau_\Gamma^y$, and $\tau^z$, which are given as the matrix elements between the LS and HS states:
\begin{align}
 \tau_\Gamma^x&=\kets{L}\bras{\Gamma} + \kets{\Gamma}\bras{L},\\
\tau_\Gamma^y&=i\kets{L}\bras{\Gamma}-i\kets{\Gamma}\bras{L},\\
\tau^z&=\sum_\Gamma\left(\kets{\Gamma}\bras{\Gamma} -\kets{L}\bras{L}\right).
\end{align}
In this representation, the $x$ and $y$ components of the PS give the local mixing between the LS and HS states and these characterize the EI state.
The $z$ component is the difference of the HS and LS densities.
Therefore, the first and second terms in Eq.~\eqref{eq:1} represent the local energy splitting between them and the Ising-type NN interaction leading to LS/HS staggered order in the case of $J_z>0$, respectively.
On the other hand, the last two terms in Eq.~\eqref{eq:1} yield the coherent quantum mixing between the LS and HS states and induce the excitonic order.
Note that, in the spin-nematic bases, the spin operators in Eq.~\eqref{eq:1} are given by
\begin{align}
 S^\Gamma=i\kets{\Gamma''}\bras{\Gamma'}-i\kets{\Gamma'}\bras{\Gamma''},\label{eq:10}
\end{align}
with $(\Gamma,\Gamma',\Gamma'')=(X,Y,Z)$ and its cyclic permutations, indicating that the spin operators mix two spin-nematic states.
The parameters in Eq.~(\ref{eq:1}) are explicitly given in the Appendix in Ref.~\cite{Nasu2016EI}.
It is worth noting that the exchange interactions between $S=1$ spins are antiferromagnetic ($J_s>0$) and those for PSs satisfy $|J_x| > |J_y|$ originating from $I>0$. Signs of both $J_x$ and $J_y$ are positive in the case of $t_b/t_a>0$, in which the noninteracting energy band in Eq~\eqref{eq:Ht} exhibits a direct gap.
Hereafter, we consider this case.

Next, we introduce the SOC.
To take into account of the $d$-orbital character of the SOC within the two-orbital model, we assume that the $a$ and $b$ orbitals represent one of the two $e_g$ orbitals and that of the three $t_{2g}$ orbitals, respectively; for simplicity, the $a$ and $b$ orbitals are identified as the $d_{x^2-y^2}$ and $d_{xy}$ orbitals, respectively, and we consider the SOC in these orbitals.
It has been confirmed in the previous studies taking account of the five $d$ orbitals that these two orbitals give dominant contributions to the EI order~\cite{PhysRevB.90.235112,PhysRevB.89.115134}.
Since the $d_{x^2-y^2}$ and $d_{xy}$ orbitals are given by the linear combinations of the states with the angular momentum $l^z=\pm 2$, $l^\pm$ vanish between these two states.
Therefore, the Hamiltonian for the SOC within these orbitals is represented as
\begin{align}
 {\cal H}_{\rm SO}=-\frac{\lambda}{\sqrt{2}} \sum_{i\eta \sigma}(l^z)_{\eta\eta'} (s^z)_{\sigma\sigma'}c_{i \eta \sigma}^\dagger c_{i \eta' \sigma'},\label{eq:soc}
\end{align}
where $l^z$ is the $l=2$ angular momentum matrix in the basis of the $d_{x^2-y^2}$ and $d_{xy}$ orbitals and $s^z$ is the $s=1/2$ spin operator.
These are given by 
\begin{align}
l^z=2
\bordermatrix{
  & {\scriptstyle a} & {\scriptstyle b}\cr
& 0 & -i\cr & i & 0
}
,\quad
s^z=\frac{1}{2}
\bordermatrix{
  & {\scriptstyle \uparrow} & {\scriptstyle \downarrow}\cr
& 1 & 0\cr & 0 & -1
}
.
\end{align}
When the pair-hopping interaction is negligibly small, the LS state can be treated as the doubly occupied state in the $b$ orbital [see Eq.~(\ref{eq:5})].
In this case, the PS operator $\tau_Z^y$ is approximately written as
\begin{align}
 \tau_{iZ}^y\sim \frac{i}{\sqrt{2}}\left(
-c_{ia\uparrow}^\dagger c_{ib\uparrow}+c_{ia\downarrow}^\dagger c_{ib\downarrow}
+c_{ib\uparrow}^\dagger c_{ia\uparrow}-c_{ib\downarrow}^\dagger c_{ia\downarrow}
\right).
\end{align}
Thus, in the low-energy subspace of ${\cal H}^{\rm eff}_{\rm el}$, the effective form of the SOC Hamiltonian is given by
\begin{align}
   {\cal H}_{\rm SO}\sim {\cal H}_{\rm SO}^{\rm eff}=-\lambda\sum_i \tau_{iZ}^y.\label{eq:7}
\end{align}

We also consider the Zeeman term caused by the magnetic field.
The total Hamiltonian is given by
\begin{align}
{\cal H}^{\rm eff}={\cal H}_{\rm el}^{\rm eff}+{\cal H}_{\rm SO}^{\rm eff}-\sum_i \bm{h}\cdot \bm{S}_i,\label{eq:6}
\end{align}
where $\bm{h}=(h^X,h^Y,h^Z)$ is the applied magnetic field.
We note that, in the present SOC, the SO(3) rotational symmetry in the triplet spin space is lost and only the rotational symmetry around the $S^Z$ axis is present even in the absence of the magnetic field.
This is because the $a$ and $b$ orbitals are identified as the $d_{xy}$ and $d_{x^2-y^2}$ orbitals, respectively, in the present model.
This point will be discussed in Sec.~\ref{sec:discussion}.

We apply the MF approximation to analyze the Hamiltonian in Eq.~(\ref{eq:6}).
In the numerical calculations, the model is defined on a two-dimensional square lattice, where the coordination number $z=4$, and the parameter values in ${\cal H}_{\rm Hubbard}$ are taken to $t_b/t_a=0.4$, $I = J$, $U/J = 6$, and $U'/J = 4$.
In the numerical calculations, the spatially uniform spin and orbital states are obtained as the MF solutions.

\section{Result}\label{sec:result}

\begin{figure}[t]
 \begin{center}
  \includegraphics[width=\columnwidth,clip]{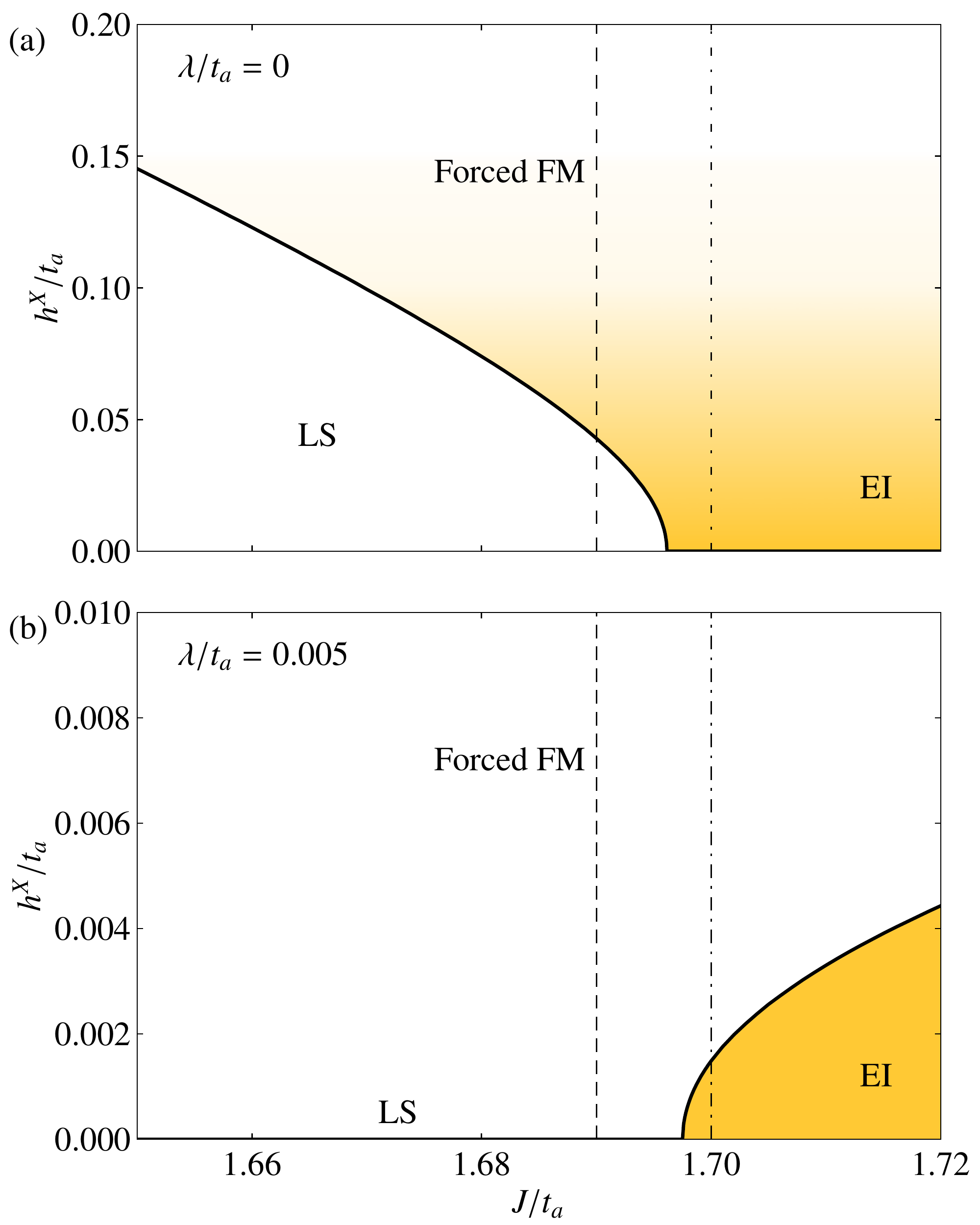}
  \caption{
Ground-state phase diagrams on the plane of the magnetic field $h^X$ and Hund coupling $J$ at (a) $\lambda=0$ and (b) $\lambda/t_a=0.005$.
The vertical dashed and dashed-dotted lines indicate the parameters for Figs.~\ref{fig_op169} and ~\ref{fig_op17}, respectively.
}
  \label{fig_phase}
 \end{center}
\end{figure}

\begin{figure}[t]
 \begin{center}
  \includegraphics[width=\columnwidth,clip]{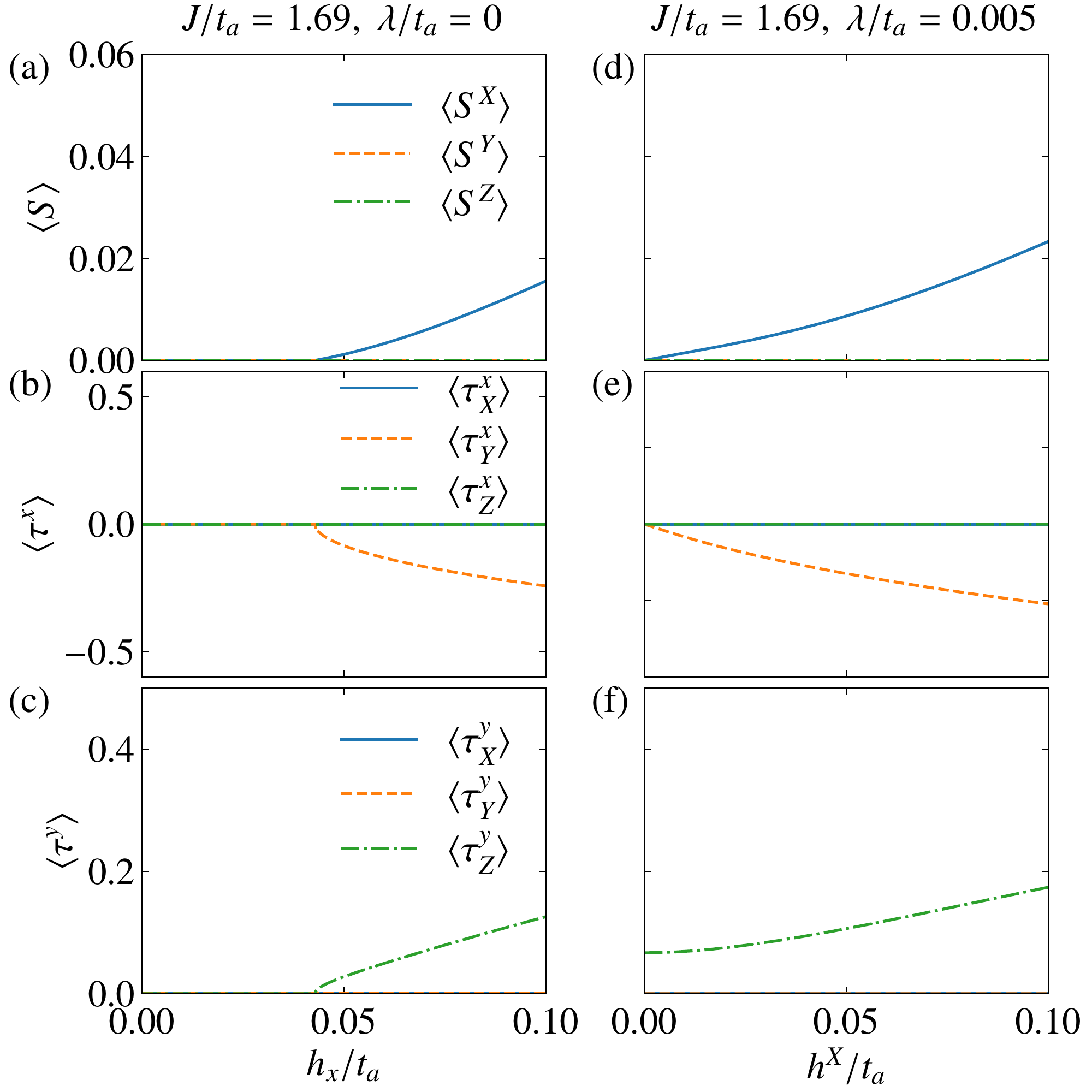}
  \caption{
Magnetic-field dependences of (a) the spin moment $\means{S^\Gamma}$ (b) PS moments $\means{\tau^x_\Gamma}$ and (c) $\means{\tau^y_\Gamma}$ at $J/t_a=1.69$ and $\lambda/t_a=0$.
(d)--(f) Corresponding figures at $\lambda/t_a=0.005$.
}
  \label{fig_op169}
 \end{center}
\end{figure}

\subsection{Phase diagram without SOC}\label{sec:wo-so}

In this section, we consider the the electronic states without the SOC.
We focus on the vicinity of the phase boundary between the LS and EI phases without magnetic orders.
This nonmagnetic EI state corresponds to the spin-triplet excitonic one, which is termed EIQ in the previous paper~\cite{Nasu2016EI}.
The ground-state phase diagram under the magnetic field without the SOC is shown in Fig.~\ref{fig_phase}(a).
In the absence of the magnetic field, the phase transition from the LS to EI state occurs at $J=J_c\simeq 1.698 t_a$ with increasing $J$.
By introducing the magnetic field, the LS state is suppressed and the EI state continuously changes into the forced FM state.
This is consistent with the previous work in Ref.~\cite{Tatsuno2016}.
Note that the phase diagram does not depend on the direction of the magnetic field because the SO(3) symmetry in the spin space exists in the absence of the magnetic field.

To discuss the field dependence of the electronic state in detail, we calculate the expectation values of the local spin moments $\means{S^\Gamma}$ and those of the PS moments, $\means{\tau_\Gamma^x}$ and $\means{\tau_\Gamma^y}$.
As mentioned before, in the present calculations, spatially uniform solutions are only obtained as the MF solution, and therefore, the site index is omitted.
Figures~\ref{fig_op169}(a)--\ref{fig_op169}(c) show these moments as functions of $h^X$ at $J/t_a=1.69$.
When $h^X$ is small, $\means{\bm{S}}=\means{\tau_\Gamma^x}=\means{\tau_\Gamma^y}=0$, indicating the LS phase ($\means{\tau^z}=-3$ is also confirmed).
Above $h_{c}^X/t_a\simeq 0.04$, the $X$ component of the spin moment becomes nonzero with accompanying the appearance of the PS moments; in the case of Fig~\ref{fig_op169}, $\means{\tau_Y^x}$ and $\means{\tau_Z^y}$ become nonzero.
This is due to the fact that the spin and PS operators are not independent of each other.
For example, the $X$ component of the spin moment is described as $S^X=-\tau_{Y}^x\tau_{Z}^y-\tau_{Z}^y\tau_{Y}^x$, which implies $\means{\tau_{Y}^x}$ is negative when $\means{\tau_{Z}^y}$ is positive under the positive magnetic field~\footnote{In general, the spin operator is given by $S^\Gamma=\left(\tau_{\Gamma'}^x \cos\theta +\tau_{\Gamma'}^y \sin\theta\right)\left(\tau_{\Gamma''}^x \sin\theta -\tau_{\Gamma''}^y \cos\theta\right)+{\rm H.C.}$ with an arbitrary value $\theta$, where $(\Gamma,\Gamma',\Gamma'')$ is the cyclic permutations of $(X,Y,Z)$.}.
We find that the spin moment $\means{S^X}$ is proportional to $(h^X-h_c^X)$ in the low-field regime, while the PS moments is to $\sqrt{h^X-h_c^X}$, being similar to the conventional order parameters in the MF theory.
This indicates that the primary order parameter of the phase transition is the PS moment.

\begin{figure}[t]
 \begin{center}
  \includegraphics[width=\columnwidth,clip]{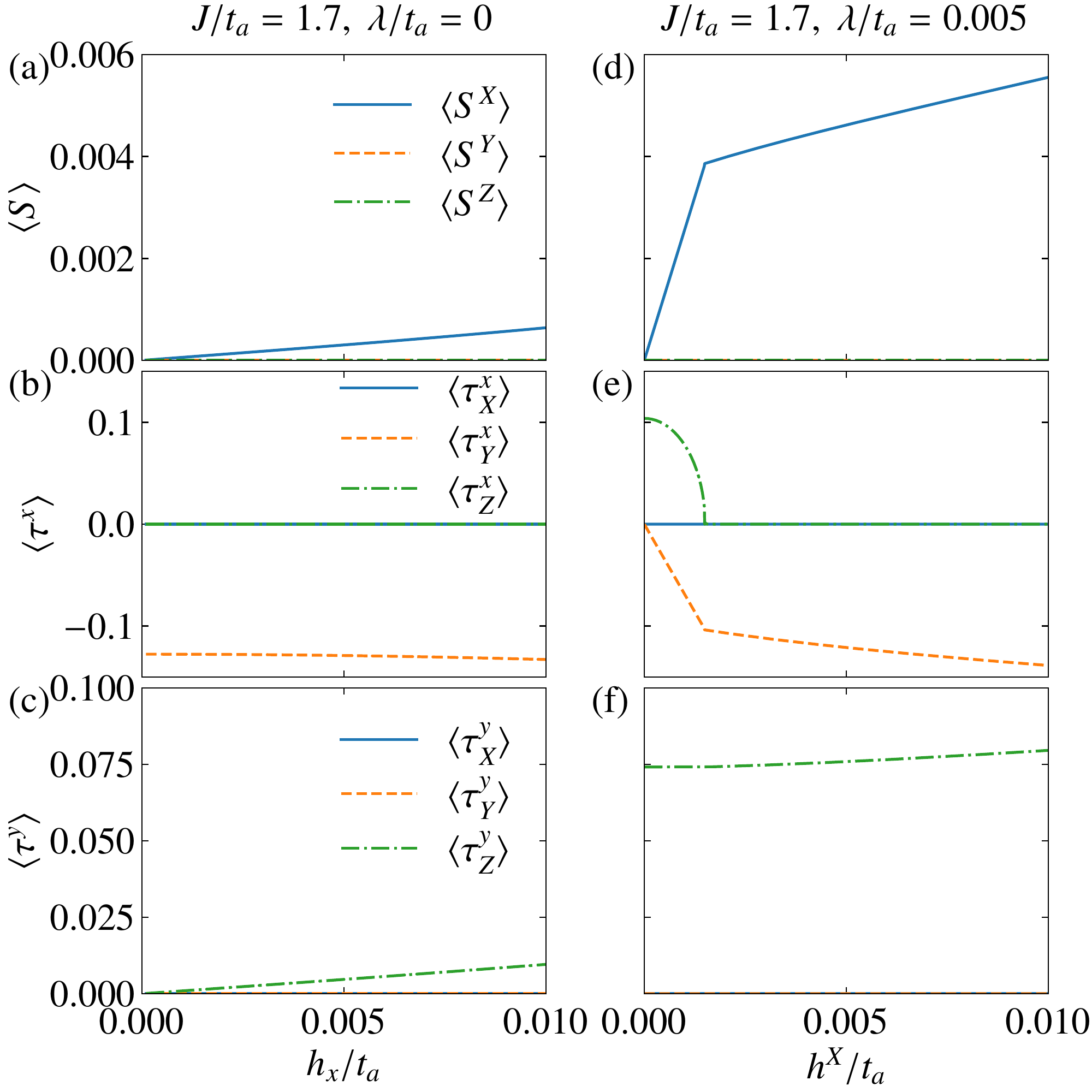}
  \caption{
Magnetic-field dependences of (a) the spin moment $\means{S^\Gamma}$ and (b) PS moments $\means{\tau^x_\Gamma}$ and (c) $\means{\tau^y_\Gamma}$ at $J/t_a=1.7$ and $\lambda/t_a=0$.
(d)--(f) Corresponding figures at $\lambda/t_a=0.005$.
}
  \label{fig_op17}
 \end{center}
\end{figure}

\begin{figure}[t]
  \begin{center}
   \includegraphics[width=\columnwidth,clip]{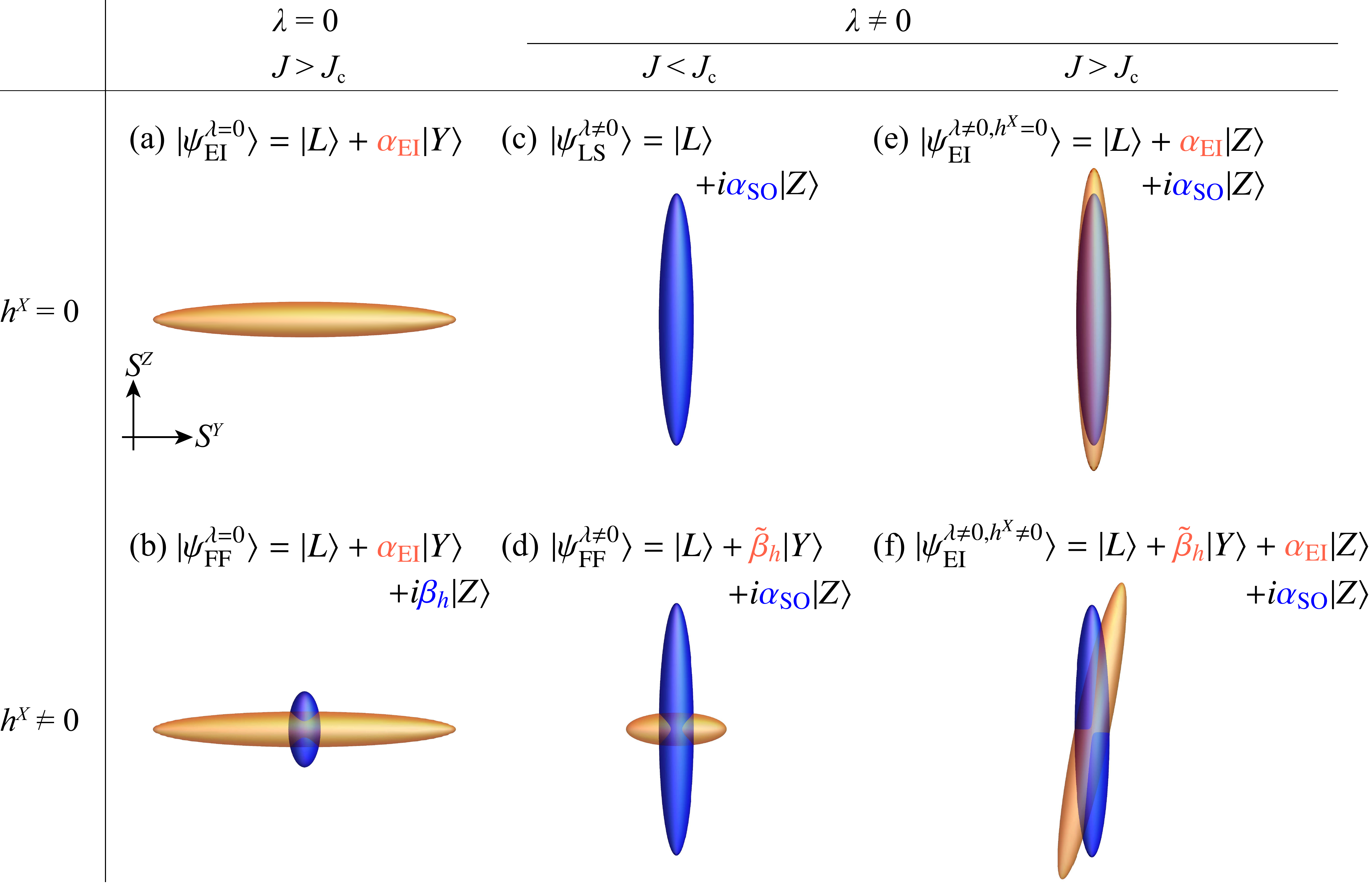}
   \caption{
 Schematic figures of the directors of the spin-nematic states on the $S^Y$-$S^Z$ plane for (a) Eq.~(\ref{eq:2}), (b) Eq.~(\ref{eq:3}), (c) Eq.~(\ref{eq:11}), (d) Eq.~(\ref{eq:4}), (e) Eq.~(\ref{eq:12}), and (f) Eq.~(\ref{eq:8}).
 The orange and blue spindles represent the real and imaginary parts of the spin nematic states when the coefficient of the LS state is taken to be unity.
 The upper panels show the states without the magnetic field and the lower ones with a small magnetic field $h^X$.
 }
   \label{fig_nematic}
  \end{center}
 \end{figure}

Figures~\ref{fig_op17}(a)--\ref{fig_op17}(c) show the expectation values of the spin and PS as functions of the magnetic field at $J/t_a=1.7$.
In the absence of the magnetic field, the EI state is realized with nonzero PS moment of $\means{\tau_Y^x}$.
With increasing $h^X$, $\means{S^X}$ and $\means{\tau_Z^y}$ linearly increase, and the EI state at $h^X=0$ is continuously connected to the forced FM state at $J/t_a=1.7$.
This result corresponds to the fact that there is no phase transition between the EI and forced FM in the phase diagram shown in Fig.~\ref{fig_phase}(a).
Further increase of the magnetic field leads to the fully spin-polarized phase composed only of the HS states, which is not included in Fig.~\ref{fig_phase}.

Here, we discuss the origin of the continuous connection between the EI and forced FM states by the magnetic field $h^X$.
At $h^X=0$, the wavefunction of the LS state is given by
\begin{align}
  \kets{\psi_{\rm LS}^{\lambda=0}}=\kets{L},
\end{align}
where the original SO(3) symmetry in the Hamiltonian is preserved.
By increasing the Hund coupling $J$, a uniform EI order is brought about with accompanying the reduction of the symmetry to U(1).
In the case of Fig.~\ref{fig_op17}(b), $\means{\tau_Y^x}$ is nonzero and the resultant U(1) symmetry is around the $S^Y$ axis, as schematically illustrated in Fig.~\ref{fig_nematic}(a).
We call this $S^Y$ the ``principal axis'' of the EI state.
The local wavefunction of the EI state is given by 
\begin{align}
  \kets{\psi_{\rm EI}^{\lambda=0}}=\kets{L}+\alpha_{\rm EI}\kets{Y},\label{eq:2}
\end{align}
where $\alpha_{\rm EI}$ is a nonzero real number. 
The hybridization of the LS state $\kets{L}$ and HS nematic state $\kets{Y}$ results in the nonzero PS moment for $\means{\tau^x_Y}$.

By introducing the magnetic field $h^X$, $\means{\tau_Z^y}$ increases proportionally to $h^X$ in the EI state similar to $\means{S^X}$, as shown in Figs.~\ref{fig_op17}(a) and \ref{fig_op17}(c).
In this case, the wavefunction is given by [see a schematic illustration in Fig.~\ref{fig_nematic}(b)]
\begin{align}
  \kets{\psi_{\rm FF}^{\lambda=0}}=\kets{L}+\alpha_{\rm EI}\kets{Y}+i\beta_{h}\kets{Z},\label{eq:3}
\end{align}
where $\beta_{h}$ is real and proportional to the magnetic field in the weak-field regime.
The coexistence of $\kets{Y}$ and $i\kets{Z}$ yields the appearance of the spin moment $\means{S^X}$.
This indicates that the EI order whose principal axis is parallel to $S^X$ is unstable under the magnetic field $h^X$ and the principal axis is selected so as to be perpendicular to $S^X$.
Therefore, the U(1) symmetry around the $S^X$ axis is absent in the presence of the EI order.
Note that, since the EI phase is originally characterized by the symmetry breaking in terms of the relative phase of the LS and HS states, the two-fold degeneracy exists for $\pm (\alpha_{\rm EI}, \beta_{h})$ in Eq.~\eqref{eq:3}.
This degeneracy disappears in the fully spin-polarized phase in the high-field regime, where the U(1) symmetry around the $S^X$ is recovered.

On the other hand, in the LS state, the magnetic field does not induce the magnetic moment up to the critical magnitude of the magnetic field because of the spin gap.
This indicates the presence of the phase transition between the LS and forced FM states at the critical field $h^X_c$, above which $\means{S^X}$ becomes nonzero as shown in Fig.~\ref{fig_op169}(b) and~\ref{fig_op169}(c).
Since the appearance of the spin moment $\means{S^X}$ requires the mixing of $\kets{Y}$ and $i\kets{Z}$ in the wavefunction, the nominal form of the wavefunction is the same as Eq.~\eqref{eq:3}.
Thus, the high-field phase above $h_c^X$ is continuously connected to the EI and forced FM phases as shown in Fig.~\ref{fig_phase}(a).
Note that the relative phase between the LS and HS states is spontaneously selected above $h_c^X$ in which a magnetic moment appears.

\subsection{Phase diagram with SOC}\label{sec:with-so}

Next, we investigate the electronic states under the SOC.
In the presence of the SOC in Eq.~\eqref{eq:6}, the SO(3) symmetry in the spin space is lowered to the U(1) symmetry around the $S^Z$ axis.
This is attributed to the fact that the $a$ and $b$ orbitals in the present model are identified as the $d_{x^2-y^2}$ and $d_{xy}$ orbitals, respectively. 
Thus, we have the following commutation relations: $[{\cal H}_{\rm SO}^{\rm eff},S_{\rm total}^X]\neq 0$, $[{\cal H}_{\rm SO}^{\rm eff},S_{\rm total}^Y]\neq 0$, and $[{\cal H}_{\rm SO}^{\rm eff},S_{\rm total}^Z]= 0$, where $S_{\rm total}^\Gamma=\sum_i S_i^\Gamma$.
Before showing the numerical results, we mention how the magnetic anisotropy appears by the SOC.
From the expression of the SOC in Eq.~(\ref{eq:7}), this stabilizes the $\kets{Z}$ ($S^Z=0$) state among the three HS states.
This is naively expected from the fact that the original SOC is diagonal for the $z$ component of spin [see Eq.~\eqref{eq:soc}].
To acquire the energy gain in the SOC, the mixing between the HS and LS states (i.e., the $d_{x^2-y^2}$ and $d_{xy}$ orbitals) is needed, where the mixed HS state is of $S^Z=0$ as the LS is the state with $S^Z=0$.
This indicates that ${\cal H}_{\rm SO}^{\rm eff}$ causes the inplane magnetic anisotropy on the $S^X$-$S^Y$ plane.
From now on, the direction of the magnetic field is chosen as $h^X$ inside of the magnetic easy plane. 

In the present study, we consider the effect of the SOC in this system, particularly for $\lambda/t_a=0.005$
(the sign of $\lambda$ does not change the phase diagram).
This is a realistic value of the magnitude of the SOC in $3d$ electron systems such as the cobaltites where $t_a\sim 1$~eV.
Although the relative value of $\lambda$ is significantly small, the physical properties are drastically changed by the introduction of the SOC as shown below.
Figure~\ref{fig_phase}(b) shows the phase diagram for $\lambda/t_a=0.005$.
The phase boundary between the EI and forced FM states is qualitatively different from that in the case without the SOC;
the EI phase is suppressed by the magnetic field, while the LS state is continuously connected to the forced FM state without any phase transitions in contrast to the result for $\lambda = 0$.

To see the magnetic-field effect in details, we show the magnetic-field dependences of the spin and PS moments in Figs.~\ref{fig_op169}(d)--\ref{fig_op169}(f) at $J/t_a=1.69$ and $\lambda/t_a=0.005$.
At $h^X=0$, $\means{\tau_Z^y}$ is nonzero due to the presence of the SOC but the spin moments $\means{S^\Gamma}$ and PS moments $\means{\tau_{\Gamma}^x}$ are zero.
This indicates that the ground state without the magnetic field is regarded as the LS state at $\lambda/t_a=0.005$ without any spontaneous symmetry breakings.
The wavefunction is given by [see a schematic illustration in Fig.~\ref{fig_nematic}(c)]
\begin{align}
  \kets{\psi_{\rm LS}^{\lambda\neq 0}}=\kets{L}+i\alpha_{\rm SO}\kets{Z},\label{eq:11}
 \end{align}
 where $\alpha_{\rm SO}$ is a nonzero real number at $\lambda\neq 0$, which results in a finite value of $\means{\tau_Z^y}$.
When $h^X$ is introduced, $\means{S^X}$ increases linearly in contrast to the case with $\lambda=0$ shown in Fig.~\ref{fig_op169}(a).
This is also understood from the finite value of the zero-temperature susceptibility due to the SOC as shown in Fig.~\ref{fig_chi_zero}, which is discussed in Sec~\ref{sec:suscep}.
In the presence of the magnetic field, the state $\kets{Y}$ is mixed to Eq.~\eqref{eq:11} so as to induce the magnetic moment $\means{S^X}$, and the resultant wavefunction is represented as  [see a schematic illustration in Fig.~\ref{fig_nematic}(d)]
\begin{align}
  \kets{\psi_{\rm FF}^{\lambda\neq 0}}=\kets{L}+i\alpha_{\rm SO}\kets{Z}+\tilde{\beta}_{h}\kets{Y},\label{eq:4}
\end{align}
where $\tilde{\beta}_h$ is a real number.
We find this wavefunction has the same form as Eq.~(\ref{eq:3}).
However, the relative phase between the LS and HS states in Eq.~(\ref{eq:4}) is fixed by $\lambda$, which is not ascribed to a spontaneous symmetry breaking, while the relative phase in Eq.~(\ref{eq:3}) is determined spontaneously by the EI order.

Next, we focus on the magnetic-field effect on the EI state at $J/t_a=1.7$ and $\lambda/t_a=0.005$.
The magnetization curves are presented in Fig.~\ref{fig_op17}(d).
The magnetization is proportional to the magnetic field $h^X$ and the magnetization curve shows a kink at $h^X=h_c^X\simeq 0.0015$.
We find that the slope below $h_c^X$ is substantially larger than that above $h_c^X$ and that at $\lambda=0$ shown in Fig.~\ref{fig_op17}(a).
Figures~\ref{fig_op17}(e) and \ref{fig_op17}(f) show the magnetic-field dependences of the PS moments.
In the absence of the magnetic field, $\means{\tau_Z^x}$ is nonzero ($\means{\tau_Z^y}$ is also nonzero due to the SOC).
With increasing the magnetic field, $\means{\tau_Z^x}$ decreases and vanishes at $h^X=h_c^X$, indicating that $\means{\tau_Z^x}$ characterizes the EI state.
Therefore, the large slope of the magnetization curve is attributed to the EI state.

Here, we discuss the reason that the EI state with nonzero $\means{\tau_Z^x}$ is stabilized in the presence of the SOC.
In the absence of the magnetic field, although the SO(3) symmetry in the spin space does not exist due to the SOC, the U(1) symmetry around the $S^Z$ axis is retained. This indicates that the uniform EI ordered state with nonzero $\means{\tau_Z^x}$ is distinguished from the uniform EI state characterized by nonzero $\means{\tau_X^x}$ and/or $\means{\tau_Y^x}$.
If the latter is realized in the absence of the magnetic field, the local wavefunction in the ground state is given by $\kets{\psi}\propto \kets{L}+c_1\kets{X}+c_2\kets{Y}+c_3\kets{Z}$, where $c_1$ and $c_2$ are real and $c_3$ is pure-imaginary due to the SOC.
In this wavefunction, from Eq.~(\ref{eq:10}), a FM order with nonzero $\means{S^X}$ and/or $\means{S^Y}$ appears without the magnetic field.
This is unfavorable to the AFM interaction in the Hamiltonian Eq.~(\ref{eq:1}).
On the other hand, in the case of the uniform EI state with $\means{\tau_Z^x}\ne 0$ and $\means{\tau_X^x}=\means{\tau_Y^x}=0$, the local wavefunction is given by a linear combination of $\kets{L}$ and $\kets{Z}$.
In this case, local spin moments do not appear, and therefore,
the uniform EI state is selected in the absence of the magnetic field.
From the above considerations, the wavefunction is uniquely given by [see a schematic illustration in Fig.~\ref{fig_nematic}(e)]
\begin{align}
 \kets{\psi_{{\rm EI}}^{\lambda\neq 0,h^X=0}}=\kets{L}+\left(\alpha_{\rm EI}+i\alpha_{\rm SO}\right)\kets{Z},\label{eq:12}
\end{align}
where the two-fold degeneracy exists for $\pm \alpha_{\rm EI}$.

By introducing $h^X$, $\means{\tau_Y^x}$ is changed proportionally to $h^X$, and $\means{\tau_Z^x}$ vanishes at $h_c^X$ as shown in Fig.~\ref{fig_op17}(e).
We identity the region with nonzero $\means{\tau_Z^x}$ as the EI phase.
In this region, the wavefunction is given by [see a schematic illustration in Fig.~\ref{fig_nematic}(f)]
\begin{align}
 \kets{\psi_{{\rm EI}}^{\lambda\neq 0,h^X\neq 0}}=\kets{L}+\left(\alpha_{\rm EI}+i\alpha_{\rm SO}\right)\kets{Z}+\tilde{\beta}_{h}\kets{Y},\label{eq:8}
\end{align}
where $\tilde{\beta}_{h}$ is a real number.
The sign of $\tilde{\beta}_{h}$ is uniquely determined under the magnetic field does not depend on that of $\alpha_{\rm EI}$.
Thus, the two-fold degeneracy originating from the EI order exits only for $\pm\alpha_{\rm EI}$.
In the case of $h^X>h_c^X$, $\means{\tau_Z^x}$ vanishes, and  the direction of the PS moment is fixed to $\means{\tau_Y^x}$, where $\alpha_{\rm EI}=0$ in Eq.~\eqref{eq:8}, which is the same as Eq.~\eqref{eq:4}.
Therefore, the phase transition from the EI to forced FM phase is understood from the flopping of the PS moment from $\means{\tau_Z^x}$ to $\means{\tau_Y^x}$ by applying $h^X$, as shown in Fig.~\ref{fig_op17}(e).

\subsection{Magnetic susceptibility}\label{sec:suscep}

\begin{figure}[t]
 \begin{center}
  \includegraphics[width=\columnwidth,clip]{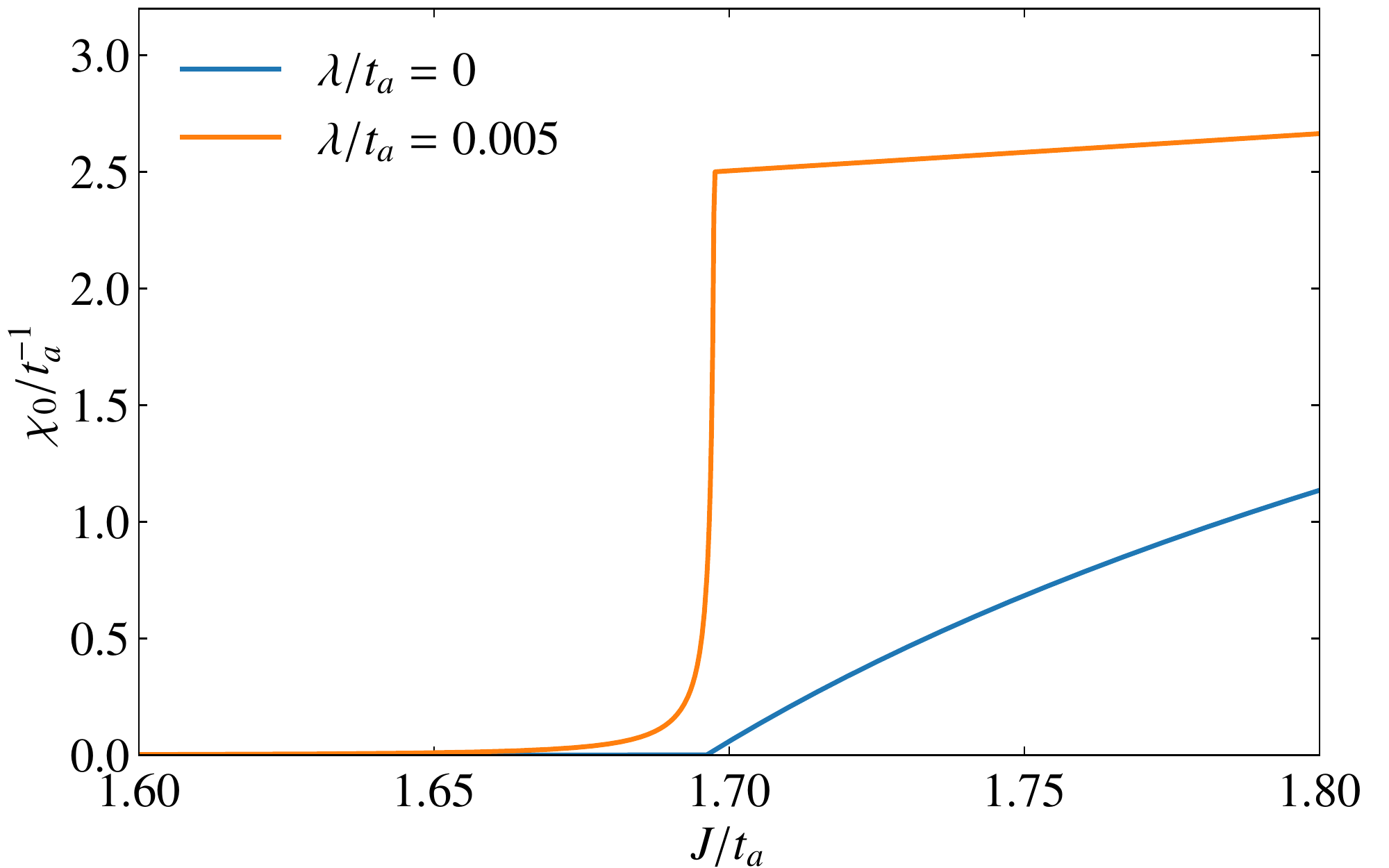}
  \caption{
Magnetic susceptibilities at zero temperature, $\chi_0$, as functions of the Hund coupling $J$.
}
  \label{fig_chi_zero}
 \end{center}
\end{figure}

The notable phenomenon caused by this PS flop is the enhancement of the slope of the magnetization in the low-field region as shown in Fig.~\ref{fig_op17}(d).
In order to show this phenomenon clearly, we calculate the magnetic susceptibility $\chi=\means{S^X}/h^x$.
We use this expression for finite-temperature calculations.
In the case of $T=0$, the calculation using $\means{S^X}/h^x$ is unstable in the vicinity of the phase boundary because the EI state is fragile under the weak magnetic field [see Figs.~\ref{fig_op17}(d)--(f)].
Instead of this approach, we compute the susceptibility at zero temperature, $\chi_0$, from the dynamical spin correlation function at $h^X=0$, which is obtained by using the spin-wave theory (see Ref.\cite{Nasu2016EI} in detail).
We have confirmed the coincidence between the results obtained from this approach and original definition except for the vicinity of the phase boundary.
Figure~\ref{fig_chi_zero} shows the $J$ dependence of $\chi_0$.
At $\lambda=0$, $\chi_0$ vanishes below $J_c (\sim 1.698t_a)$ because of the spin gap in the LS state.
In the EI state realized above $J_c$, $\chi_0$ increases from zero at $J_c$ with increasing $J$.
In the presence of the SOC, $\chi_0$ changes continuously but is strongly enhanced at around $J_c$ as shown in Fig.~\ref{fig_chi_zero} for $\lambda/t_a=0.005$.

\begin{figure}[t]
 \begin{center}
  \includegraphics[width=\columnwidth,clip]{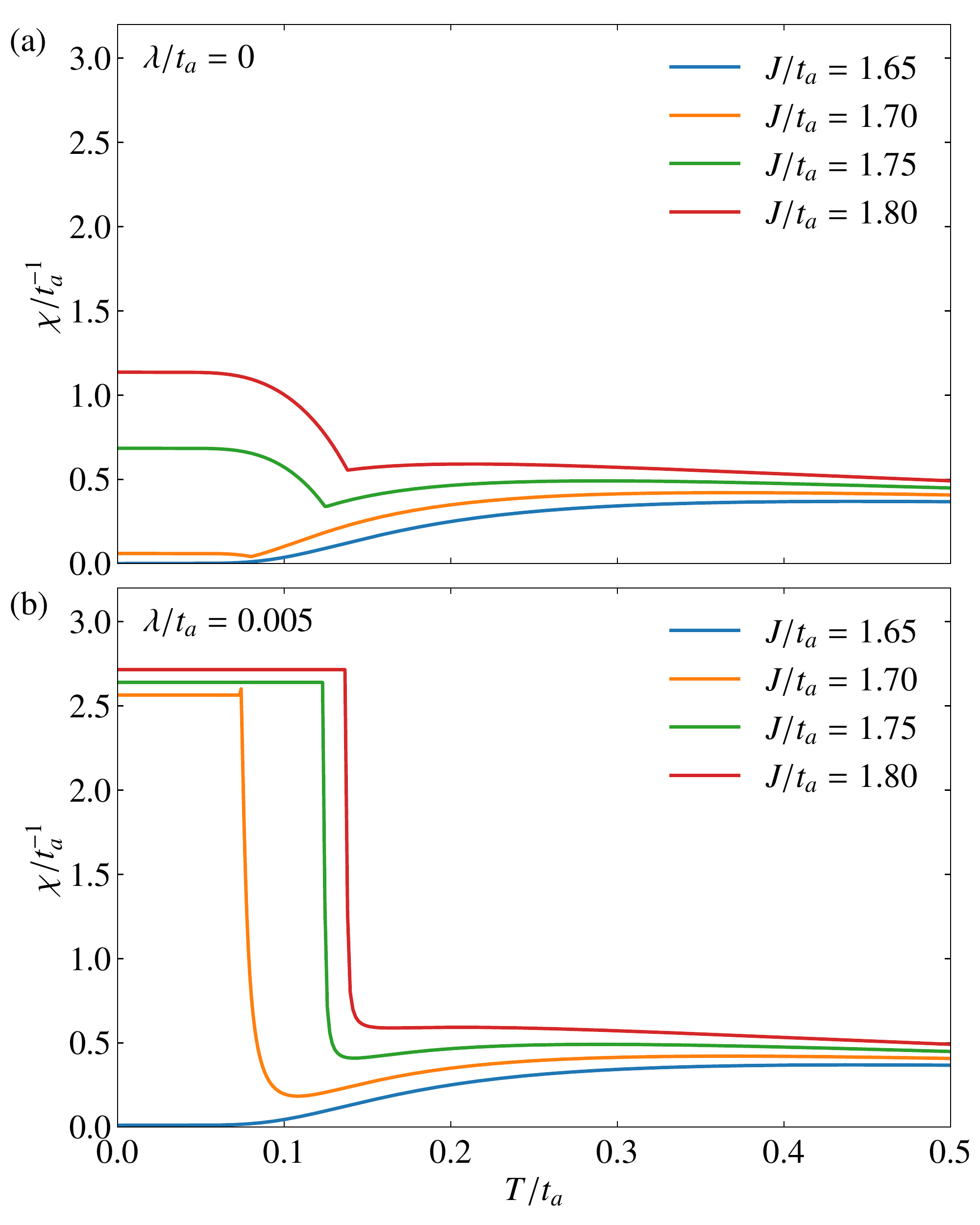}
  \caption{
Magnetic susceptibilities as functions of temperature at (a) $\lambda/t_a=0$ and (b) $\lambda/t_a=0.005$ for the several values of the Hund coupling.
}
  \label{fig_chi}
 \end{center}
\end{figure}

This enhancement is also observed in the results for the temperature dependence.
Figures~\ref{fig_chi}(a) and \ref{fig_chi}(b) show the finite-temperature susceptibility $\chi=\means{S^X}/h^x\rvert_{h^X\to 0}$ calculated by the MF approximation at $\lambda/t_a=0$ and $0.005$, respectively.
First, we focus on the case at $\lambda/t_a=0$.
In high temperatures, the nonzero susceptibility is observed since the HS states are thermally excited.
In the case of $J/t_a=1.65$, the LS ground state is continuously connected to the high-temperature paramagnetic state.
On the other hand, above $J_c$, we find the phase transition at a certain temperature to the low-temperature EI phase, which is accompanied by the enhancement of the susceptibility.
Next, we discuss the results in the case of the nonzero SOC at $\lambda/t_a=0.005$.
As shown in Fig.~\ref{fig_chi}(b), the phase transition is observed with the saturation of the susceptibility at $J/t_a=1.70$, $1.75$, and $1.80$ but is not at $J/t_a=1.65$.
While the finite-temperature phase transition occurs in the same manner with the zero SOC case, the strong enhancement is observed slightly above the transition temperature.
Below the transition temperature, $\chi$ is almost constant as a function of temperature.
This behavior is in contrast to the case at $\lambda=0$, where $\chi$ continues to change with decreasing temperature from the critical temperature.

\subsection{Origin of enhancement of susceptibility}\label{sec:fluc}

\begin{figure}[t]
 \begin{center}
  \includegraphics[width=\columnwidth,clip]{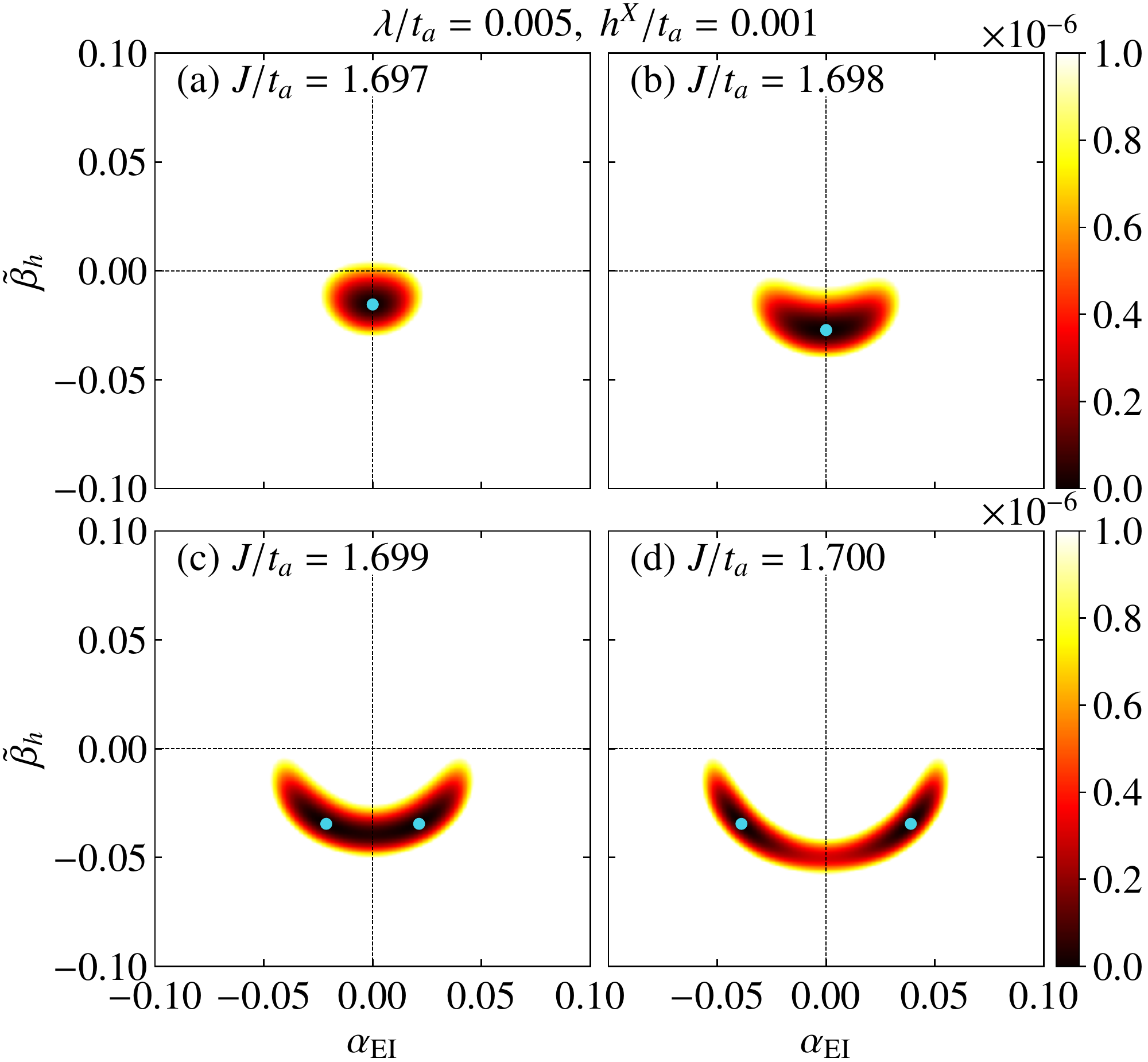}
  \caption{
    Adiabatic energy planes as functions of the parameters $\alpha_{\rm EI}$ and $\tilde{\beta}_h$ in the wavefunction Eq.~(\ref{eq:8}) for (a) $J/t_a=1.697$, (b) $J/t_a=1.698$, (c) $J/t_a=1.699$, and (d) $J/t_a=1.7$ with $(\lambda, h^X)/t_a=(0.005, 0.001)$.
    Blue points correspond to the MF solutions.
   }
  \label{fig_adiabatic}
 \end{center}
\end{figure}

\begin{figure}[t]
  \begin{center}
   \includegraphics[width=\columnwidth,clip]{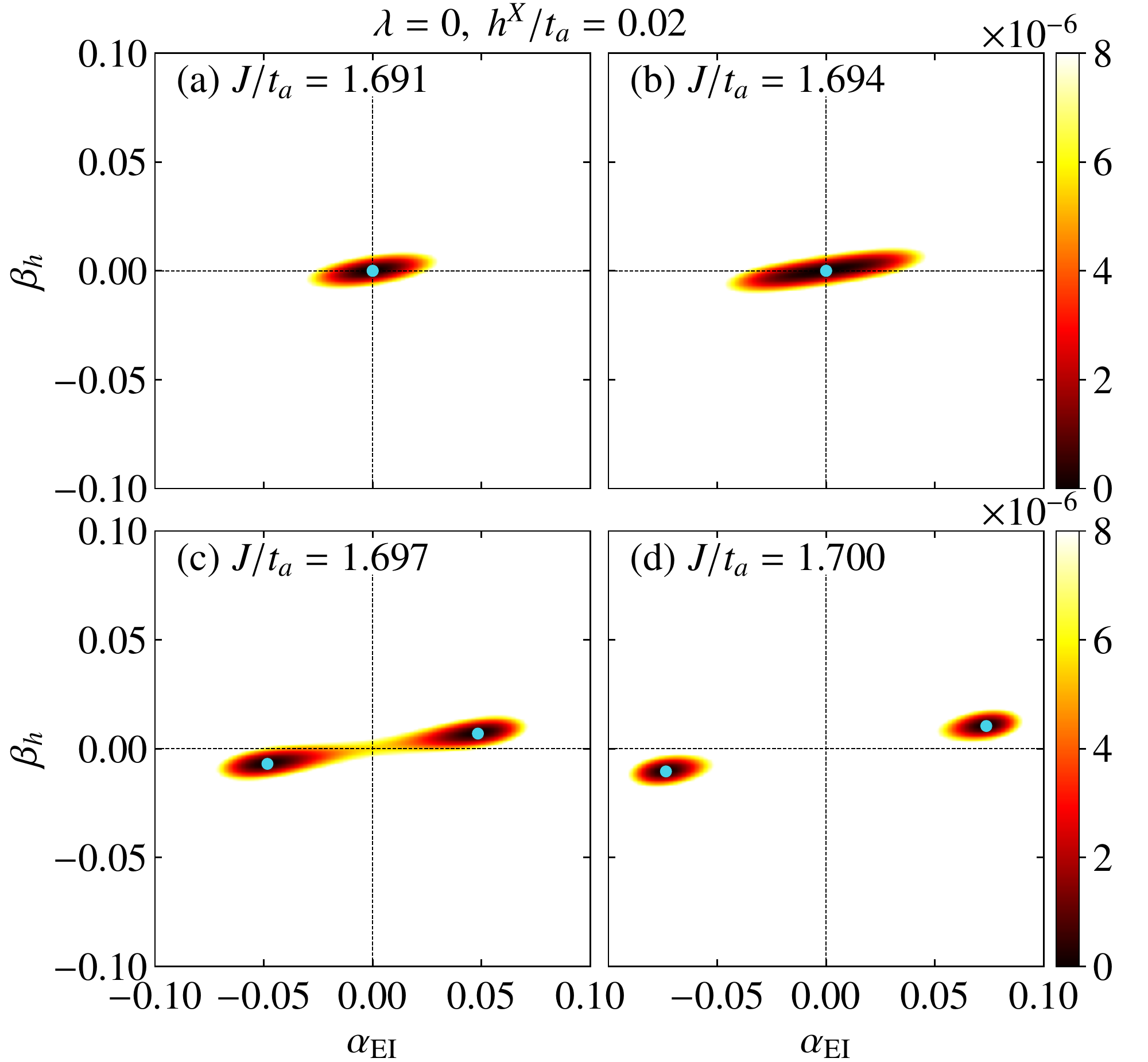}
   \caption{
    Adiabatic energy planes as functions of the parameters $\alpha_{\rm EI}$ and $\beta_h$ in the wavefunction Eq.~(\ref{eq:3}) for (a) $J/t_a=1.691$, (b) $J/t_a=1.694$, (c) $J/t_a=1.697$, and (d) $J/t_a=1.7$ with $(\lambda,h^X)/t_a=(0, 0.02)$.
    Blue points correspond to the MF solutions.
    Note that $\beta_h$ is the imaginary part of the coefficient for $\kets{Z}$ in Eq.~\eqref{eq:3}.
 }
   \label{fig_adiabatic2}
  \end{center}
 \end{figure}

Here we discuss the origin of the enhancement of the susceptibility in the presence of the SOC.
The susceptibility is the response of $\means{S^X}$ to $h^X$.
In this case, the magnetic field yields the HS state $\kets{Y}$ with the coefficient $\tilde{\beta}_h$ as discussed in Eq.~\eqref{eq:8}.
In the EI phase, the coefficient of $\kets{Z}$, i.e., $\alpha_{\rm EI}$, is also nonzero, and therefore, we expect the presence of the fluctuation between the spin-nematic states $\kets{Y}$ and $\kets{Z}$.
To examine the nematic fluctuation, we calculate the adiabatic energy in the plane of $\alpha_{\rm EI}$ and $\tilde{\beta}_h$.
The adiabatic energy is defined by the expectation value of the MF Hamiltonian for the wavefunction where these coefficients are regarded as variables under the fixed MFs giving the ground-state MF solution.
Figure~\ref{fig_adiabatic} shows the adiabatic energy plane in the vicinity of $J_c$ at $\lambda/t_a=0.005$ in the presence of the week magnetic field $h^X/t_a=0.001$.
While only one minimum exists below $J_c$, two minima at nonzero $\alpha_{\rm EI}$ are found above $J_c$ as a consequence of the EI order associated with the spontaneous Z$_2$ symmetry breaking.
A notable point is that the arc-shape low-energy region exists between these minima. 
This suggests the presence of the nematic fluctuation between $\kets{Y}$ and $\kets{Z}$ caused by the rotational mode.
Owing to this fluctuation, $\tilde{\beta}_h$ is easily changed by the magnetic field $h^X$, and as the result, the large susceptibility is observed in the EI state in the presence of the SOC.

On the other hand, in the case of $\lambda=0$, the wavefunction in the presence of $h^X$ is given in Eq.~\eqref{eq:3}, where $\beta_{h}$ is the imaginary part of the coefficient for $\kets{Z}$.
The adiabatic energy planes are shown in Fig.~\ref{fig_adiabatic2}.
In the LS phase [Figs.~\ref{fig_adiabatic2}(a) and~\ref{fig_adiabatic2}(b)], the minimum is unique at $(\alpha_{\rm EI},\beta_h)=(0,0)$, corresponding to $\kets{L}$.
In the EI phase, two minima are found at $\pm (\alpha_{\rm EI},\beta_h)$ because of the $Z_2$ degeneracy in the symmetry broken states.
The minima are apart from each other, and therefore, the nematic fluctuation is weaker than that at $\lambda\neq 0$.

\section{Discussion and Summary}\label{sec:discussion}

Here, we discuss the relevance to real materials such as cobaltites.
In the present study, we address the two-orbital model, where the $d_{x^2-y^2}$ and $d_{xy}$ orbitals are taken into account.
The selection of $d_{x^2-y^2}$ and $d_{xy}$ orbitals among the five $d$ orbitals gives rise to the symmetry lowering in the real space, namely, the $z$ axis is inequivalent to the other axes.
This suggests that a structural transition or an enhancement of structural distortion should occur with the phase transition to the EI phase.
Indeed, the increase of the distortion of CoO$_6$ octahedra has been found by the neutron diffraction measurement at $T_S\sim$90K in Pr$_{0.5}$Ca$_{0.5}$CoO$_{3}$~\cite{Tong2009,PhysRevB.66.052418,doi:10.1143/JPSJ.73.1987}, which supports the selection of the two orbitals, in addition to the first-principles calculations~\cite{PhysRevB.90.235112,PhysRevB.89.115134,Yamaguchi2017}.
The $d_{x^2-y^2}$ and $d_{xy}$ orbitals are composed of the $l^z=\pm 2$ states, indicating that the large SOC is expected for the EI state composed of the two orbitals.
Thus, we expect that an abrupt change of the magnetization is induced by the magnetic field.
This corresponds to a large magnetic susceptibility, which will provide another piece of evidence for the EI state in cobaltites.

Moreover, the collective excitations from the EI state have been theoretically proposed~\cite{Nasu2016EI,Yamaguchi2017,Geffroy2019} and should be observed by the inelastic neutron scattering.
In the present study, we find the enhancement of the low-energy fluctuations of the spin-nematic states in the vicinity of the phase transition to the EI phase.
We expect that the fluctuations appear as a gapped mode due to the SOC.
Indeed, a low-energy gapped excitation has been observed by the inelastic neutron scattering~\cite{Moyoshi2018}.
This might correspond to the fluctuation of the spin nematicity of the HS or IS states but the detailed relationship remains a future issue.
Futhermore, optical measurements is also a promising route to reveal the low-energy fluctuations of spin nematicity in the EI state~\cite{Murakami2017,Murakami2020,murakami2020pre,andrich2020pre}.

In summary, we have investigated the effect of the SOC on the EI state by analyzing the low-energy effective Hamiltonian.
We find that magnetic susceptibility is strongly enhanced in the vicinity of the phase transition from the LS to EI states in the presence of the SOC.
This originates from the fluctuation of the spin nematicity intrinsic in the $S=1$ HS states.
The present study not only offers the way to identify the EI state through the significant change of the experimentally accessible quantity but also will stimulate further investigations for clarifying the role of the spin nematicity in the EI state.

\begin{acknowledgments}
This work is supported by Grant-in-Aid for Scientific Research under Grant No.~JP16K17747, JP17H02916, JP18H05208, JP19K03723, and JP20H00121.
Parts of the numerical calculations were performed in the supercomputing systems in ISSP, the University of Tokyo.
\end{acknowledgments}

\bibliography{refs}

\end{document}